\documentclass[twocolumn,prl,superscriptaddress,showpacs,preprintnumbers,amsmath,amssymb]{revtex4}

\usepackage{graphicx}
\usepackage{dcolumn}
\usepackage{bm}
\usepackage{color}

\newcommand{\mct}{{\mbox{\scriptsize mct}}}
\newcommand{\ini}{{\mbox{\scriptsize eq}}}
\newcommand{\fit}{{\mbox{\scriptsize (fit)}}}
\newcommand{\maxx}{{\mbox{\scriptsize (max)}}}
\newcommand{\theory}{{\mbox{\scriptsize (the)}}}
\newcommand{\IS}{{\mbox{\tiny IS}}}
\newcommand{\phiJ}{\varphi_{\mbox{\scriptsize J}}}
\newcommand{\phiini}{\varphi_{\mbox{\ini}}}
\newcommand{\phimct}{\varphi_{\mbox{\scriptsize mct}}^{\theory}}
\newcommand{\phimctfit}{\varphi_{\mbox{\scriptsize mct}}^{\fit}}

\begin{document}
\title{Jamming Transition and Inherent Structures of Hard Spheres and Disks}
\author{Misaki Ozawa}
\affiliation{Institute of Physics, University of Tsukuba, Tennodai 1-1-1, Tsukuba 305-8571, Japan}
\author{Takeshi Kuroiwa}
\affiliation{Institute of Physics, University of Tsukuba, Tennodai 1-1-1, Tsukuba 305-8571, Japan}
\author{Atsushi Ikeda}
\thanks{Current address: Laboratoire Charles Coulomb, UMR 5221 CNRS, Montpellier, France}
\author{Kunimasa Miyazaki}
\affiliation{Institute of Physics, University of Tsukuba, Tennodai 1-1-1, Tsukuba 305-8571, Japan}

\date{\today}
\begin{abstract}
Recent studies show that volume fractions $\phiJ$ at the jamming
 transition of frictionless hard spheres and disks are not uniquely determined but exist 
over a continuous range.
Motivated by this observation, we numerically investigate the dependence of 
$\phiJ$ on the initial configurations of the parent fluid equilibrated 
at a 
volume fraction $\phiini$,  before compressing to generate a jammed packing.
We find that $\phiJ$ remains constant when $\phiini$ is small but
sharply increases as $\phiini$ exceeds the dynamic transition point 
which the mode-coupling theory predicts. 
We carefully analyze configurational properties of both jammed packings
and parent fluids and find that, while all jammed
packings remain isostatic, the increase of $\phiJ$ is 
accompanied with subtle but distinct changes of local orders, 
a static length scale, and an exponent of the finite size scaling. 
These results  are consistent with the scenario of the random first order transition theory
of the glass transition. 
\end{abstract}
 \maketitle

Despite of their apparent similarities, a unifying theory of the
glass transition of supercooled fluids and the jamming transition of
athermal particles such as granular materials is still missing. 
Both are characterized by a transition from a flowing state to 
a randomly jammed state at a finite density or temperature.  
The glass transition is achieved by cooling equilibrium fluids
slowly  (but quickly enough to avoid crystallization), whereas  a common 
protocol to induce the jamming transition is to compress dilute hard
sphere/disk systems rapidly.
For frictionless particle systems (which we shall consider in this Letter), 
it has long been argued that the jamming transition 
is interpreted as the zero-temperature limit of the glass
transition~\cite{liu1998b}.
Numerical studies, however, show that these two transitions are distinct
and their natures are more complicated~\cite{Ikeda2012}.
For example, the jamming transition has been believed to take place
sharply at a unique volume fraction in the thermodynamic limit, the so-called ``points J''; 
$\phiJ\approx 64\%$ for three dimension (3D)
and $84\%$ for two dimension (2D)~\cite{O'Hern2003}. 
But recently it has been demonstrated that $\phiJ$ is not unique but
exists over a continuous range of volume fractions whose values 
vary depending on the protocols used to generate the jammed
states~\cite{Chaudhuri2010,Ciamarra2010d,Schreck2011}; 
$\phiJ$ becomes larger than $64\%$ or $84\%$ if 
one prepares moderately dense systems or thermally
equilibrated systems at low temperatures and then rapidly compresses 
to generate the jammed states. 
Surprisingly, the jammed configurations at different $\phiJ$
are found to remain
isostatic, lack a partial crystalline order, and 
therefore are not mixtures of ordered and ``maximally
random jammed'' (MRJ) states~\cite{Schreck2011,Torquato2000}. 

On the other hand, our understanding of the glass transition 
is no better than that of the jamming transition. 
Even a {\em mean field picture} of the glass transition has not been
established. 
A promising candidate 
is the so-called random first order transition (RFOT) theory, originally
inspired by the mean-field theory of spin-glasses~\cite{kirkpatrick1989,Biroli2009}.  
Crudely speaking, RFOT integrates the energy landscape picture, the concept of the
ideal glass transition, and the
mode-coupling theory (MCT)~\cite{goldstein1969,Stillinger1982,Gotze2009}. 
Despite its theoretical coherence, this RFOT-MCT scenario still 
remains controversial, partly due to the lack of impeccable numerical
and experimental evidence. 

The goal of this Letter is to provide numerical evidence that 
the protocol dependence of $\phiJ$ is a natural consequence of the
RFOT-MCT scenario and thus the scenario can unify the glass and jamming
transitions of frictionless particles. 
The idea that the 
energy landscape of glasses is intimately
related to the jamming transition is not
new~\cite{Mari2009,Parisi2010,Chaudhuri2010,Charbonneau2011c}.   
But to the best of our knowledge, quantitative characterizations of 
the transition points, particle configurations, and the associated length
scales 
have not been done so far. 
First, we shall briefly recapitulate the essence of the RFOT-MCT scenario~\cite{Biroli2009} 
and how it relates the jamming to the glass transition~\cite{Mari2009}.   
In the mean-field limit, RFOT predicts that, 
as a fluid is cooled down, it first undergoes the dynamical transition
at a temperature $T_{\mct}$ followed by the thermodynamic transition at
a lower temperature $T_K$.
Below $T_{\mct}$,  the multidimensional 
energy surface becomes suddenly rugged.
The energies at local minima of the surface or the inherent
structures (IS), $e_{\IS}$, which are almost constant at high
temperatures start decreasing at $T_{\mct}$.
Concomitantly, the saddles of the 
energy surface vanish and all stationary points become stable.
Dynamics near $T_{\mct}$ is described by MCT.
It predicts that
dynamical quantities such as the relaxation time $\tau_{\alpha}$
diverge with a  power law $|T-T_{\mct}|^{-\gamma}$
where $\gamma$ is a parameter also calculated by MCT~\cite{Gotze2009}.
In finite dimensions, however, the dynamic transition is smeared out by 
activation hoppings between local minima separated by finite barriers
and becomes merely a crossover.
An important observation is that the geometrical properties of the energy
landscape are not controlled by a single temperature $T_{\mct}$ any more. 
Simulations have revealed that $e_{\IS}$ starts decreasing abruptly 
at a onset temperature $T_o$, whereas saddles survive well below $T_o$ 
until they vanish at $T_{th}$, a so-called threshold
temperature~\cite{sastry1998,angelani2000}.  
On the other hand, the relaxation time obtained by simulations is still well fitted
by MCT's power law,  $\tau_\alpha \sim |T-T_{\mct}^{\fit}|^{-\gamma}$,
but $T_{\mct}^{\fit}$ used for fitting was found to be considerably
lower than $T_{\mct}^{\theory}$, the value obtained theoretically by solving the MCT equation~\cite{Gotze2009}. 
Surprisingly, $T_{\mct}^{\fit}$ turned out to be very close to
$T_{th}$~\cite{angelani2000}, whereas  
$T_{\mct}^{\theory}$ is close to $T_o$~\cite{sastry1998,Brumer2004b}. 
Discrepancies between $T_{\mct}^{\fit}$ $(\approx T_{th})$ and
$T_{\mct}^{\theory}(\approx T_{o})$, both of which should be identical in the
mean-field limit, are due to the non-mean-field effect and 
can be explained using kinetic arguments~\cite{Mayer2006b,Bhattacharyya2008}.  

The above argument also applies to hard sphere fluids. 
The temperature and energy $(T, e_{\IS})$, relevant variables for continuous potential
fluids, should be replaced by the (inverse) pressure $P^{-1}$ and volume
for hard-core potential systems~\cite{Stillinger1964}. 
Instead of the volume, we shall adopt the density, or volume fraction $\varphi$.
The inherent structures $\varphi_{\IS}$ are obtained by 
compressing a parent fluid equilibrated at a finite $P$ by letting 
$P\rightarrow\infty$ (with an extra minimization using a conjugate gradient method), 
just as $T$ is quenched to zero to obtain $e_{\IS}$ for continuous potential fluids.
This is nothing less than a process to generate jammed packings for
frictionless hard spheres and thus $\varphi_{\IS}$ should be equivalent
with $\phiJ$.
Employing the RFOT scenario discussed above, we predict that
$\varphi_{\IS}$ or $\phiJ$ is unchanged as long as $P$ of the parent fluid equilibrated at  
a volume fraction
$\phiini$ 
is low but starts {\it increasing} as $P$ (or $\phiini$) exceeds $P_{\mct}$ (or  $\varphi_{\mct}$).
In other words, $\phiJ$ is not a unique value but 
is a function of $P$ or $\phiini$ and can exist over a continuous
range~\cite{Mari2009,Parisi2010,Charbonneau2011c}.    
The largest $\phiJ^{\maxx}$ would correspond to the inherent
structures of the fluid at the thermodynamic transition point, {\em i.e.}, $\phiini=\varphi_{K}$.
For finite dimensional systems, 
$\phimct$ obtained from MCT theoretically should be lower than
$\phimctfit$ obtained by fitting the simulation data for 
the relaxation times. 
Furthermore, $\phiJ$ should increase if we prepare a dense parent fluid 
such that $\phiini > \phimct$ (rather than $\phimctfit$).
Another important prediction of RFOT is that, at the dynamic
transition point, the system enters to the coexisting region of numerous
metastable phases, or {\em mosaics}. 
Thus,  the {\em static} length scale associated with the mosaics, if
any, should appear at $\varphi_{\mct}^{\theory}$.
There have been several attempts to directly measure the static length
in supercooled fluids but most studies have focused on 
the configurations of parent fluids and at far lower temperatures or higher
densities than the dynamic transition points~\cite{Biroli2009}.

In order to verify these predictions, we prepare 
thermally equilibrated hard spheres (3D) and disks (2D) at
various initial fractions $\phiini$ and study their inherent structures $\phiJ$.
Both systems studied here are 50:50 binary mixtures with a size ratio of 1.4
with periodic boundary conditions~\cite{O'Hern2003,Chaudhuri2010}.
The systems are equilibrated at $\varphi_{\ini}$ using Monte-Carlo
simulation and then compressed rapidly to generate jammed states.
Following the procedure employed in
Refs.~\cite{O'Hern2003,Desmond2009b}, we switch the hard-core potential
with the soft harmonic potential just before the compression, allowing
particles to overlap.
The system is then relaxed to the zero-energy state using the conjugate
gradient method. 
This compression and energy-minimization cycle is iterated till 
the volume fraction is maximized without particle overlap.
Note that the algorithm to generate the jammed states is
essential. 
For example, the Lubachevsky-Stillinger algorithm is inappropriate
because the system 
keeps equilibrating during the slow compression and finds lower local minima
of the landscape or higher $\phiJ$~\cite{Lubachevsky1990}.
\begin{figure}[t]
\begin{center}
\includegraphics[width=1\columnwidth]{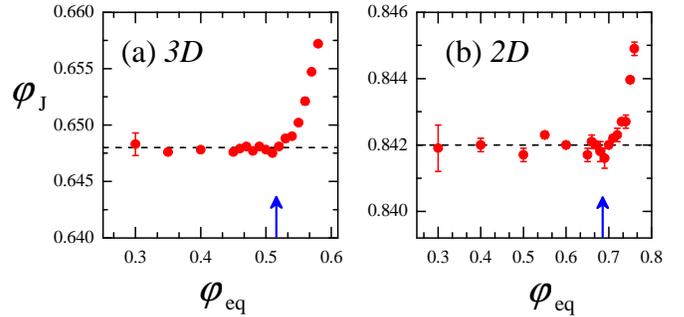}
\caption{$\phiJ$ as a function of $\varphi_{\ini}$ for
binary mixtures in (a) 3D and (b) 2D.  Arrows indicate the positions of
 $\phimct$.
Broken horizontal lines are $\phiJ$ reported by O'Hern {\em et al.}~\cite{O'Hern2003}.
} 
\vspace*{-0.6cm}
\label{figure1}
\end{center}
\end{figure}
The system sizes are varied from $N=64$ to $2048$. 
$\phiJ$ in the
large $N$-limit is evaluated using the finite size scaling
$|\phiJ(N)-\phiJ| \sim N^{-1/\nu d}$, where $d$ is the spatial dimension.
In Figure 1, the dependence of $\phiJ$ on $\varphi_{\ini}$ is shown.
The exponents $\nu=0.72$ (3D) and 0.74 (2D) obtained for the
smallest $\phiini$ are used for the rest of data.  
Actually, we found that $\nu$ varies noticeably depending on
$\varphi_{\ini}$ as we shall discuss  below. 
However, the different $\nu$'s do not affect appreciably the results of 
Fig.\ref{figure1}, other than more scattering of data points and larger error
bars. 
At small $\varphi_{\ini}$, the jamming transition points are identical
with those already reported in the literatures, $\phiJ \approx 0.648$ (3D)
and $0.842$ (2D)~\cite{O'Hern2003}. 
However, $\phiJ$ abruptly starts increasing at large $\phiini$.
The onset fractions are found to be very close to
$\phimct$ independently evaluated by solving the MCT equations for 
binary mixtures using the static structure factor matrix  $S(k)$ 
obtained by the Percus-Yevick theory  as an input. 
Indicated by arrows in Fig.\ref{figure1} are 
$\phimct\approx 0.516$ (3D) and $0.685$ (2D). 
We confirmed that these values do not vary more than 2\% if
simulated $S(k)$ is used.
The onset points are obviously much lower than $\phimctfit\approx 0.59$ (3D)
and $0.79$ (2D) obtained from fitting the relaxation data~\cite{Brambilla2009,Weysser2011b}. 
Results shown in Fig.1 are consistent with those reported in Refs.~\cite{Chaudhuri2010, Schreck2011}.
\begin{figure}[thb]
\begin{center}
\includegraphics[width=0.95\columnwidth]{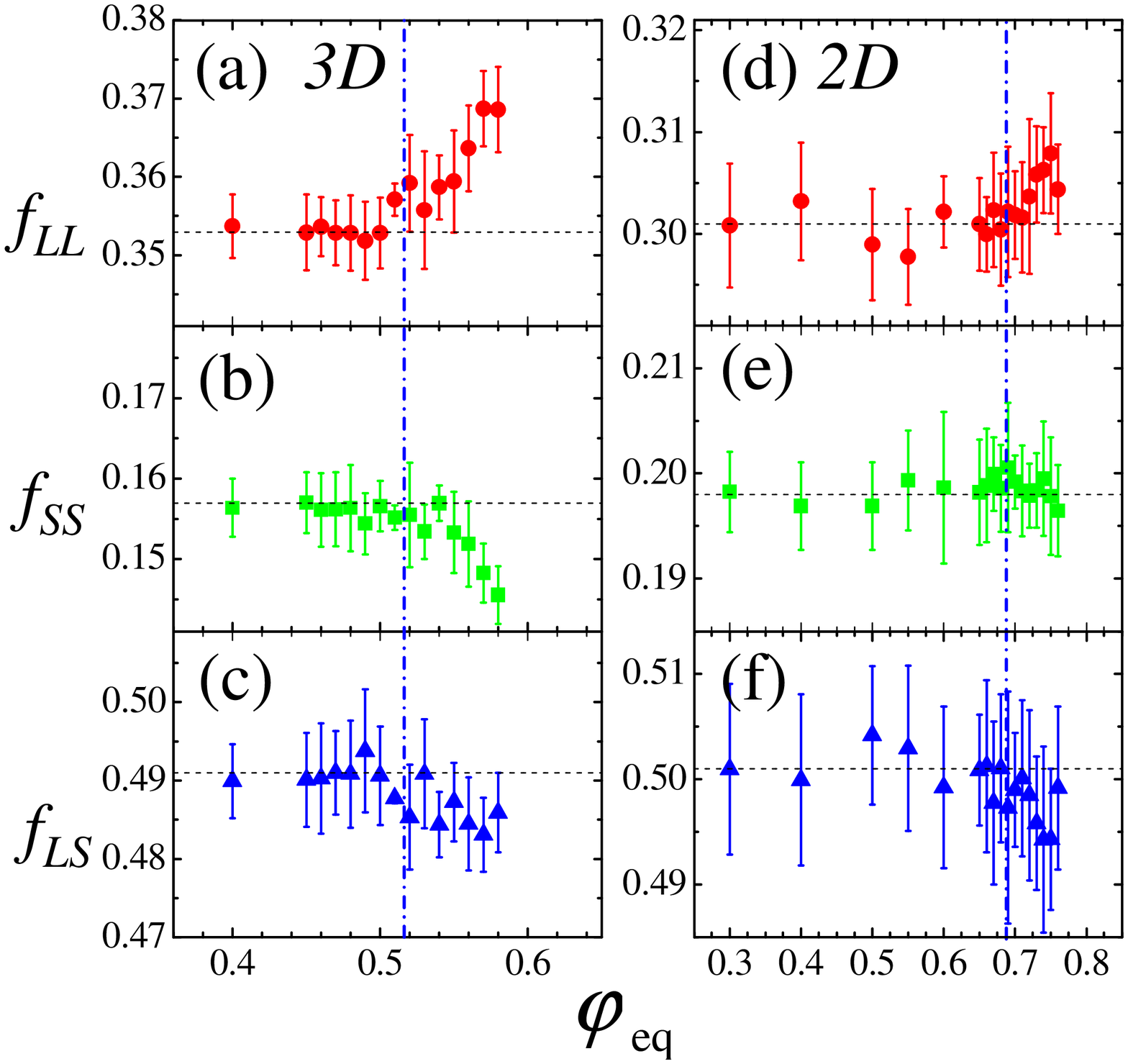}
\caption{Compositional order parameters 
for 3D (a)$\sim$(c) and 2D (d)$\sim$(f). Vertical
 dash-dot lines represent $\phimct$ and horizontal broken lines are
 guides for eyes. }
\vspace*{-0.6cm}
\label{figure2}
\end{center}
\end{figure}
We also found that, as $\phiJ$ increases,  the jammed configurations 
remain isostatic, {\em i.e.}, the contact number at
$\phiJ$ is given by $z=2d$, whereas the number of rattlers slightly increases~\cite{Chaudhuri2010}. 
We also measured the time sequence of the inherent structures for
several $\phiini$ and observed that the patterns of 
sequences qualitatively change from white-noise-like at 
$\phiini <\phimct$  to step-wise at $\phiini > \phimct$ (not shown),
implying that the nature of the landscape is altered~\cite{denny2003}.  
These results support quantitatively that the jamming and glass
transitions can be discussed under the common rubric of the RFOT-MCT
scenario and also that $\phimct$ is not a fictitious value of an approximate theory 
but bears the essential geometrical meaning. 

In order to clarify the nature of the denser jammed packings obtained
from the parent fluid at $\phiini > \phimct$, we focus on properties of their configurations. 
We calculate the compositional and orientational orders.
Figure \ref{figure2} shows the dependence on $\phiini$ 
of the compositional order parameters of the jammed packings $(f_{LL}, f_{SS}, f_{SL})$, 
the number fractions of the contact pairs of the {large (L)}  and {small
(S)}  particles~\cite{Schreck2011}.  
\begin{figure}[thb]
\begin{center}
\includegraphics[width=0.9\columnwidth]{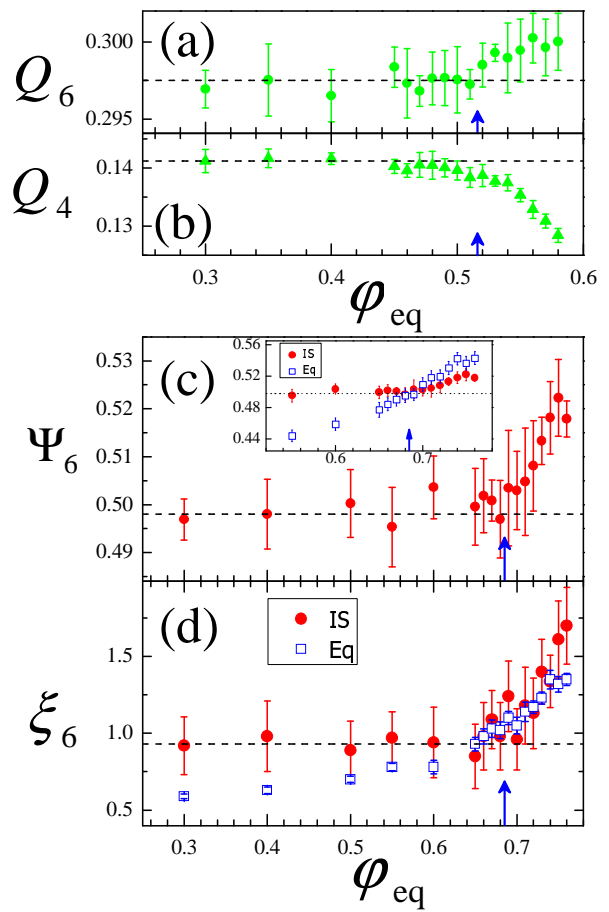}
\caption{BOO parameters for large particles of
jammed packings,  (a) $Q_6$  and (b) $Q_4$ for 3D and (c) $\Psi_6$ for 2D. 
The inset in (c) is $\Psi_6$ for the parent  fluid (Eq; empty squares)
and the jammed packing (IS; filled  circles). 
(d) The correlation length $\xi_6$ obtained from $g_6 (r)$ of the jammed packing 
and the parent fluid for 2D.
Horizontal broken lines are guides for eyes. 
}
\vspace*{-0.6cm}
\label{figure3}
\end{center}
\end{figure}
For ideally random configurations, $f_{LL}+f_{SS}\approx f_{SL}\approx
0.5$ holds. 
Though this is the case for all $\phiini$,  
minute but sharp increase of $f_{LL}$ and decrease of $f_{SS}$ are observed
at $\phiini \approx  \phimct$.
For 3D, the variations are about 
5\%. 
Qualitatively similar changes are observed for 2D, consistent with Ref.~\cite{Schreck2011}.
We next analyze the bond-orientational order (BOO) parameters $Q_4$ and
$Q_6$ (3D) and
$\Psi_6$ (2D) defined in Refs.~\cite{Schreck2011,Steinhardt1983}. 
The BOO parameters evaluated for the large particles are shown in Figure \ref{figure3} (a)$\sim$(c).  
The results for the small particles 
show qualitatively similar behavior although the variations are less pronounced.
All results demonstrate that the BOO parameters are constant at $\phiini< \phimct$
but change abruptly at $\phimct$.
One may want to argue that the synchronized change of $\phiJ$ and
the compositional/orientational orders is due to the onset of a partial 
crystallization or demixing and that the system traces a line connecting smoothly 
the MRJ packing at the smallest $\phiJ$ and the ideally ordered
configuration at the maximal density~\cite{Torquato2000}. 
If it is the case, however, the isostaticity should break down 
and variations of $f_{ij}$ and the BOO parameters would be far larger than those shown in Figs.\ref{figure2} and
\ref{figure3}~\cite{Schreck2011}. 
Of course, we did not observe any sign of demixing from the eye inspection of the jammed configurations.
These facts strongly suggest that the system is riding on a different branch.
We emphasize that these sharp changes at $\phimct$ are only observed for
the jammed packings.
The inset of Fig.\ref{figure3} (c) shows that $\Psi_6$ of the parent fluid
continuously increases with $\phiini$ with no hint to change around $\phimct$.
Similar results were obtained for 3D.  

According to RFOT, the increase of $\phiJ$ should be
accompanied with the appearance of numerous metastable states or mosaics and 
the system ``phase-separates'' into these states.
Thus, it is expected that the mosaics and their associated length scale
should appear at $\phimct$.
To detect a hint of the emergence of such states, we calculate the static correlation function of the
fluctuations of the local BOO parameters 
$g_6(r)=\langle \delta\Psi_6({\bf r})\delta \Psi_6(0)\rangle$ for 2D
and extract out the length scale $\xi_6$ by fitting the results with the Ornstein-Zernike
function (Fig.~\ref{figure3} (d)). 
\begin{figure}[thb]
\begin{center}
\includegraphics[width=1\columnwidth]{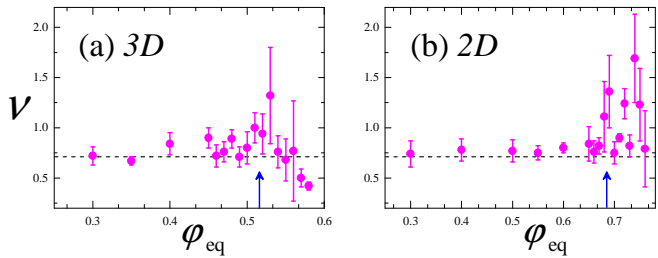}
\caption{$\phiini$-dependence of $\nu$ of the finite-size scaling law 
$|\phiJ(N)-\phiJ| \sim N^{-1/\nu d}$ for (a) 3D  and (b) 2D. 
The horizontal broken lines are values reported in Ref.~\cite{O'Hern2003}.}
\label{figure4}
\end{center}
\vspace*{-0.6cm}
\end{figure}
Similar result was obtained for 3D.
$\xi_6$ which is constant at low $\phiini$ starts increasing
at $\phimct$. 
Also shown is $\xi_6$ obtained for the parent fluid, which  
monotonically increases with $\phiini$.
The sudden increase of $\xi_6$ at $\phimct$ for the jammed packing, 
which is not observed for the parent fluid, suggests a possibility that it is a direct reflection of 
the emergence of the mosaics. 

Finally, we argue that $\phimct$ may also mark the point beyond which 
the finite-size scaling law is qualitatively altered due to the emergence
of mosaics. 
In the crossover region at which the MCT's
critical dynamics and activation hoppings coexist, the finite size effect is highly
nontrivial according to the RFOT-MCT scenario~\cite{Berthier2012}.  
For the short-range interaction systems, 
these two mechanisms may compete and a simple power-law scaling may be
violated. 
Figure \ref{figure4} shows $\phiini$-dependence of the finite size
scaling exponent $\nu$, obtained by naively using the scaling law.
$\nu$'s are constant at $\phiini < \phimct$ and close to the values reported
in Ref.~\cite{O'Hern2003} but 
start fluctuating and become errant at $\phiini\approx\phimct$.  
We presume that this is another, though indirect, evidence supporting the
RFOT-MCT scenario. 

In summary, we have accumulated and displayed quantitative evidence 
that the RFOT-MCT scenario integrates the jamming and glass transitions in a
common language and successfully explains the continuous increase of $\phiJ$ reported previously.
We demonstrated for the first time that the dynamical transition point
$\phimct$ theoretically evaluated, and not  $\phimctfit$ obtained by the
fitting, unambiguously marks the onset of qualitative changes of 
the energy landscape, or the ``volume landscape'' for hard spheres/disks.
Note that the results shown here are consistent with those for various
short-ranged potential systems~\cite{Brumer2004b} but not for the fully-connected models~\cite{Mari2011b}. 
In Ref.~\cite{Mari2011b}, the onset volume fraction at which $\phiJ$
starts increasing is considerably smaller than $\phimct$ obtained from
the simulated relaxation time in the mean-field regime.
This contradictory result might be due to the long-ranged interaction of
the model.
Indeed, it is known that the onset temperature of the inherent structures 
for a fully-connected spin-glass model of a finite size is much higher than the 
mean-field value and the convergence to the mean-field limit is
extremely slow~\cite{Crisanti2000d}.

All results in this Letter eloquently support the RFOT-MCT scenario but many nagging questions are left for us. 
For example, 
{\em why does the MCT work quantitatively so well in finite dimensions?} 
It is especially puzzling because recent studies show that the
traditional MCT is not perfectly consistent with the mean-field scenario
at large spatial dimensions~\cite{Ikeda2010}.  
Also we are left unanswered about the relation of the static length which
we observed with other lengths observed via static and dynamic
measurements in the past~\cite{Biroli2009,Mosayebi2010,Tanaka2011}.  
And the last interesting question may be whether the configurational
properties, especially isostatic nature, of jammed packings are affected when $\phiini$ 
exceeds $\phimctfit$ at which all saddles of the 
energy surface near the IS vanishes. 
These are a few of many problems which are left for the future works.

\acknowledgments
This work is supported by the JSPS Core-to-Core Program ``International research network for
non-equilibrium dynamics of soft matter'', KAKENHI No. 2154016, 24340098, and
Priority Areas ``Soft Matter Physics''. 
We thank the Research Center for
Computational Science, Okazaki and ISSP of Tokyo University for
the use of supercomputers.
We wish to thank L. Berthier, P. Charbonneau, S. Sastry, and F. Zamponi for valuable
discussions.

\end{document}